# A Trust Based Fuzzy Algorithm for Congestion Control in Wireless Multimedia Sensor Networks (TFCC)

Arpita Chakraborty[1], Srinjoy Ganguly[2], Mrinal Kanti Naskar[3], Anupam Karmakar[4]

[1] Dept. of Electronics and Communication Engineering, Techno India, Salt Lake, Kolkata 700091, India
[2, 3] Dept. of Electronics and Telecommunication Engineering, ADES Lab, Jadavpur University, Kolkata 700032, India
[4] University of Calcutta, Dept. of Electronic Science, Kolkata 700009, India
[1]carpi.technoindia@yahoo.com, [2]srinjoy_ganguly92@hotmail.com, [3]mrinalnaskar@yahoo.co.in, [4]anupamkarmakar@yahoo.co.in

*Abstract -* **Network congestion has become a critical issue for resource constrained Wireless Sensor Networks (WSNs), especially for Wireless Multimedia Sensor Networks (WMSNs) where large volume of multimedia data is transmitted through the network. If the traffic load is greater than the available capacity of the sensor network, congestion occurs and it causes buffer overflow, packet drop, deterioration of network throughput and quality of service (QoS). Again, the faulty nodes of the network also aggravate congestion by diffusing useless packets or retransmitting the same packet several times. This results in the wastage of energy and decrease in network lifetime. To address this challenge, a new congestion control algorithm is proposed in which the faulty nodes are identified and blocked from data communication by using the concept of trust. The trust metric of all the nodes in the WMSN is derived by using a two-stage Fuzzy inferencing scheme. The traffic flow from source to sink is optimized by implementing the Link State Routing Protocol. The congestion of the sensor nodes is controlled by regulating the rate of traffic flow on the basis of the priority of the traffic. Finally we compare our protocol with other existing congestion control protocols to show the merit of the work.**

## I. INTRODUCTION

In recent years Wireless Multimedia Sensor Networks (WMSNs) are being developed for multimedia applications like video surveillance, video traffic control systems, health monitoring and industrial process control. WMSNs are a special kind of Wireless Sensor Networks (WSNs), in which sensor nodes contain low cost CMOS cameras and microphones which are capable to retrieve multimedia data like still images, voice, audio and video streams as well as scalar sensor data from the physical environment [1]. Similar to WSNs, WMSNs consist of one or more sink nodes or base stations and several hundreds of source nodes, deployed randomly in an area. The multimedia content of WMSNs are of two types**:** The snapshot type multimedia data which is obtained from event triggered observations in a short time period and streaming multimedia content which is generated over a longer time period with continuous information delivery [2]. When the heavy traffic load of multimedia data from source to sink is greater than the available capability of the network, congestion occurs in the upstream direction. The resulting effect is buffer overflow, increased packet latency, packet drop, deterioration of network throughput and quality of service (QoS). Again, the sensor networks are prone to various types of security attacks due to the presence of faulty or malicious nodes. The HELLO flood attack, jamming attack, Sybil attack or node replication attack aggravate network congestion many folds by diffusing useless packets or retransmitting the same packet several times. It causes wastage of energy and decrease in network lifetime. Also, it hampers the event detection reliability of the sensor network significantly. Therefore, congestion avoidance (CA) or congestion detection (CD), and congestion control (CC) are important issues in WMSNs.

Unfortunately, the traditional congestion control protocols do not consider the role of malicious nodes. We propose a novel trust based Fuzzy congestion controller for WSNs/ WMSNs. In the proposed model, the misbehaviour of the sensor nodes is identified using the concept of trust. The malicious nodes are thus isolated and blocked from the data communication.Technically, Trust modeling is a mathematical representation of a node's opinion about another node in the network [3]. Trust value of each node w.r.t another node is evaluated dynamically on the basis of some parameters known as Trust Metrics (TM) which measure the node's behavior during previous data transfer through this node and recommendations received from other trusted nodes [4]. Different TMs can be used for trust estimation of the sensor nodes. Some commonly used TMs are data packet forwarding rate, data packet precision, data packet delay, availability based on beacon/hello messages, routing protocol execution, consistency of reported values, reputation, sensing communication, packet address modification and remaining battery life [5]. In the proposed work, the trust of an individual sensor node with respect to all of its one-hop neighbours is computed from TMs using a fuzzy inferencing scheme. If the trust value of a node of a particular one hop neighboring node exceeds a predefined Trust Threshold ($T_{TH}$), then the node is referred to benevolent node or trusted node with respect to that particular node [5]. The link between these two nodes are called trusted link. Data packet transmissions are possible only through the trusted link. The same node may not be act as trusted node with respect to the other one hop neighboring node. In this case, trust value of the node on that particular neighboring node is below $T_{TH}$ and the link between these two nodes are untrusted link. Data packets could not be transmitted through the untrusted link. The nodes without any trusted link are referred to as malicious or faulty nodes. All the malicious nodes are blocked from the network so that they could not take part in data packet routing.

In next step, the congestion level of each trusted node is computed from the buffer queue length of the node. The congestion-trust metric is formed using the proposed inference engine. The available routes from source to sink are obtained by applying the Link State Routing Protocol (LSRP). A Fuzzy control mechanism is proposed to prevent congestion of the sensor nodes by regulating the rate of traffic flow on the basis of the priority of the traffic source.

The Fuzzy Logic Controller (FLC) used in the present work basically consists of four components : Fuzzifier, Fuzzy IF-THEN rules, Fuzzy Inference mechanism and Defuzzifier. The Fuzzifier converts crisp input data to suitable input Fuzzy sets. Fuzzy Inference mechanism derives Fuzzy output by combining Fuzzy rules into a mapping routine from input to output of the system. Finally Defuzzifier converts the output fuzzy set to a crisp value. The general FLC block diagram is shown in Fig. 1

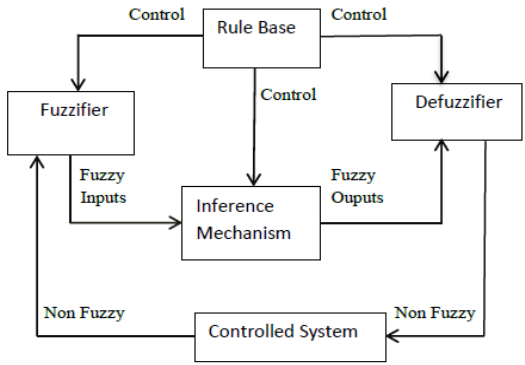

Fig. 1 Generalized Fuzzy Logic

The rest of the paper is organized as follows**:** In section II, some existing congestion control protocols, related to the proposed work are discussed. In section III, we present our proposed work, Trust Based Fuzzy Algorithm for Congestion Control in WMSNs (TFCC). The simulation results and comparison with other existing protocols is given in section IV and finally section V concludes the paper.

## II. RELATED WORKS

Trust based congestion control in WSNs and WMSNs is a new research topic and has not been discussed in great depth. Most of the existing algorithms related to congestion of WSNs are based on the traditional approach of congestion detection and control. In CODA [6], Wan et al. propose a congestion detection and avoidance technique for WSNs, using an open loop hop by hop back pressure and closed loop multisource regulation. Congestion is detected by periodical sampling the channel load and current buffer occupancy. Hull et al. integrated three complimentary congestion control strategies in [7], where different layers of the traditional protocol stack: hop by hop flow control, rate limiting and a prioritized MAC protocol are taken care of. In ESRT [8], both congestion detection and reliability level are estimated at the sink. The major limitations of [8] are to support only the single hop operation and pushing all complexity to the sink. In PSFQ [9], slow down distribution and quick error recovery method are considered to avoid congestion. The drawback of PSFQ includes several timers setting, highly specialized parameter tuning and complicated internal operations. In PCCP [10], hop by hop congestion control is imposed on the basis of congestion degree as well as node priority index which is introduced to reflect the importance of each node. QCCP-PS [2] uses a queue based congestion indicator and can adjust the source traffic rate based on current congestion in the upstream nodes and the priority of each traffic source. QCCP-PS shows some improvement over PCCP. However, both PCCP and QCCP-PS do not consider existence of faulty nodes in the network. In FCC protocol [11], Zarei et al. propose a Fuzzy based trust estimation for congestion control in WSNs. The trust concept is used for detection of malfunctioning nodes using fuzzy logic. The resulting effects are: a decrease in the packet drop ratio and accordingly an increase in the packet delivery rate. FCCTF protocol[12] is basically a modification of FCC, in which Threshold Trust Value is used for decision making.. Threshold could change dynamically with increasing or decreasing traffic of the related region. Although faulty nodes are isolated in FCC and FCCTF, still there is further scope of improvement in terms of congestion detection and congestion regulation considering traffic priority.

## III. PROPOSED WORK

The growing popularity of multimedia applications in WSNs has led to its increased utilization and hence congestion in the network. Furthermore, inexpensive sensor nodes are prone to failure and various security attacks, when deployed randomly in remote and harsh area. The faulty nodes may drop the received packets or re-forward it several times. Thus diffusion of useless packets by the faulty nodes may increase network

### A. Step One

In this step, the misbehavior of the sensor nodes is detected using the concept of trust. The trust value of each individual sensor node is estimated by using a Fuzzy algorithm. The Trust Threshold ($T_{TH}$) value is assigned, which is application dependent. As $T_{TH}$ value is increased, the network security is also increased. The malicious nodes are isolated and blocked for further data packet transmission and finally we apply Link State Routing Protocol for getting the available routes. The pictorial depiction of step one of our algorithm is shown in Fig.2

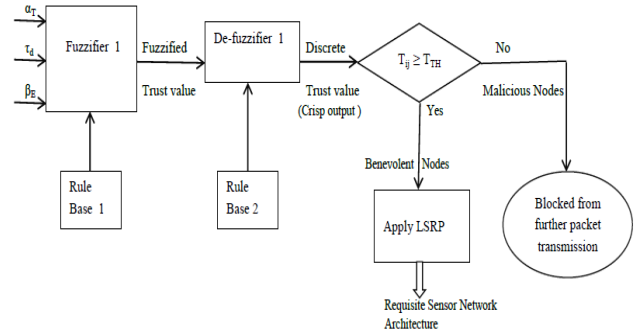

Fig. 2  Flow diagram  of TFCC (step one)

*1)  Computation of Trust Value :*

Any number of Trust Metrics (TM) can be taken for the computation of trust in the sensor nodes. In this model, three important TMs, namely Packet Transmission Ratio ($\alpha_T$), Packet Latency Ratio ($\tau_d$) and Remaining Energy Ratio ($\beta_E$) have been taken as the crisp input variable at Fuzzifier 1.  The Fuzzified Trust value of the sensor nodes are obtained at the output using Rule Base 1 (consisting of 64 rules).

The Packet Transmission Ratio is defined as the ratio of the number of acknowledgements received ($\alpha_r$) from the suspected node to the total number of packets ($\alpha_{tx}$) sent to the suspected node. Mathematically, $\alpha_T = \alpha_r / \alpha_{tx}$ where $\alpha_r \leq \alpha_{tx}$ and $0 < \alpha_T < 1$. The case $\alpha_r > \alpha_{tx}$ implies the misbehavior of the node in which the suspected node transmits useless packets or retransmits the same packets several times. The Packet Latency Ratio is the latency of the suspected node ($\tau_s$) to the mean latency of the nodes other than the suspected node ($\tau_{av}$). Mathematically, $\tau_d = \tau_s / \tau_{av}$. The case, $\tau_s > \tau_{av}$ implies that the latency of the suspected node is more that the average value and it is considered as the misbehavior of the node. Again, if the input packets are sent very fast without data aggregation, the latency in the suspected node would be very low, which implies $\tau_s \ll \tau_{av}$ and it is also considered as the misbehavior of the node.

The Remaining Energy Ratio ($\beta_E$) is defined as the ratio of the present battery voltage ($\beta_p$) to the full battery voltage ($\beta_F$) of the sensor node.  Mathematically, $\beta_E = \beta_P / \beta_F$ and $0 < \beta_E < 1$
The Table I shows the crisp input ranges of the Fuzzy variable at the Fuzzifier 1.

TABLE I
Crisp Input Range And Fuzzy Variable At Fuzzifier 1

| Trust metric | Crisp input range | Fuzzy variable range |
|---|---|---|
| Packet Transmission Ratio, $\alpha_T$ | 0 - 0.45 | VL ( Very Low ) |
| | 0.4 – 0.6 | L  ( Low) |
| | 0.55 – 0.75 | M (Medium) |
| | 0.7 - 1 | H (High) |
| Packet Latency Ratio, $\tau_d$ | 0 – 0.45 | VLD ( Very Low Delay ) |
| | 0.4 – 0.6 | LD ( Low Delay) |
| | 0.55 – 1 | AD ( Average Delay) |
| | Greater than 1 | HD ( High Delay ) |
| Remaining Energy Ratio, $\beta_E$ | 0 – 0.45 | VLE ( Very Low Energy ) |
| | 0.4 – 0.6 | LE (  Low Energy ) |
| | 0.55 – 0.75 | ME(Medium Energy ) |
| | 0.7 - 1 | HE ( High Energy ) |

The Table II as shown below, describes the Rule Base 1 (comprising of 64 rules), which is implemented in Fuzzifier 1.

TABLE II
Rule Base 1

| $\alpha_T$ | $\tau_d$ | $\beta_E$ | Fuzzy Trust Value |
|---|---|---|---|
| VL/L/M/H | VLD/LD/AD/HD | VLE/LE | VLT |
| VL/L | VLD/HD | ME | VLT |
| VL/L | VLD/HD | HE | VLT |
| VL/L | AD/LD | ME | MT |
| VL/L | AD/LD | HE | MT |
| M/H | AD/LD | ME | HT |
| M/H | AD/LD | HE | HT |
| M/H | VLD/HD | ME | LT |
| M/H | VLD/HD | HE | LT |

The Fuzzy trust value  of the nodes are categorically divided into four classes such as Very Low Trust (VLT), Low Trust (LT), Medium Trust (MT) and High Trust (HT) respectively. The crisp trust value of each node is obtained from Defuzzifier 1, controlled by rule base 2 as shown in Table III below.

TABLE III
Rule Base 2

| Fuzzy variable range for Trust | Crisp Trust Value, $T_{ij}$ |
|---|---|
| VLT | 0 – 0.45 |
| LT | 0.4 – 0.6 |
| MT | 0.55 – 0.75 |
| HT | 0.7 – 1 |

## 2) Segregation of Malicious Nodes :

Let us consider $T_{ij}$, $T_{ik}$, and $T_{ip}$ be the trust value of node i with respect to node j, node k and node p respectively. All the nodes denoted by j, k and p are the one hop neighbors of $i^{th}$ node. If $T_{ij} \geq T_{TH}$, the $i^{th}$ node is called the benevolent or trusted node with respect to node j and the link between node i and j is known as the trusted link. Again, if $T_{ik} < T_{TH}$, the node i would not be considered as the benevolent node with respect to node k, although it behaves as benevolent with respect to node j. The link between node i and k is considered as the untrusted link. Similarly, if $T_{ip} \geq T_{TH}$, the link between node i and node p is trusted link. For $T_{ip} < T_{TH}$, it is untrusted link. Data packet routing is possible only through the trusted link. If a node does not have any trusted link with its one hop neighbor, it is called the malicious node. Hence, the malicious nodes are identified from the misbehavior of the nodes and the value of $T_{TH}$. In our model, $T_{TH} = 0.5$ (this value is application specific[4] and may vary as per the requirement). The malicious nodes of the sensor network are blocked and would not be able to take part in data packet transmission and data routing.

## 3) Link State Routing Protocol ( LSRP) :

In the proposed model, LSRP is applied on the trusted links of the benevolent nodes to get the available routes in the upstream direction from source to sink. The route search algorithm used in the proposed model is described in our previous work FTRSP [13].

## B. Step Two

Here, the congestion of each trusted node is estimated by a Fuzzy algorithm considering the current buffer queue size of the corresponding sensor node. The parameter Complementary Congestion Index (CCI) is calculated to represent the congestion status of the nodes. Two fixed threshold $C_{TH}(Min)$ and $C_{TH}(Max)$ (values defined in the range of the queue length). If the queue length of a node is less than $C_{TH}(Min)$, congestion is low; when it is in-between $C_{TH}(Min)$ and $C_{TH}(Max)$, the congestion is medium and related linearly to the queue length; The congestion is high when queue length is more than $C_{TH}(Max)$ [2]. Next, a congestion-trust metric is generated and finally, a Fuzzy trust-based congestion control scheme is proposed in which the rate of incoming traffic flow of the sensor node is adjusted dynamically on the basis of the priority of the traffic. The Fig. 3 shows the flow diagram of step two of the proposed model.

## 1) Computation of Complementary Congestion Index :

The congestion status of a trusted node is obtained from Complementary Congestion Index (CCI) which is a function of buffer queue length. It is assumed that each sensor node has more than one buffer and separate queue to store input packets from each child nodes and from its own local source. So, if $i^{th}$ node has $N_i$ child nodes, then the total number of queues in $i^{th}$ node is given by $N_i +1$ [2]. For $k^{th}$ queue of $i^{th}$ node, the queue size is denoted by $Q_s(k)$ and we define a parameter $I_K'(i) = f(Q_s(k))$ where $I_K'(i)$ is the CCI of $i^{th}$ node for $k^{th}$ queue.

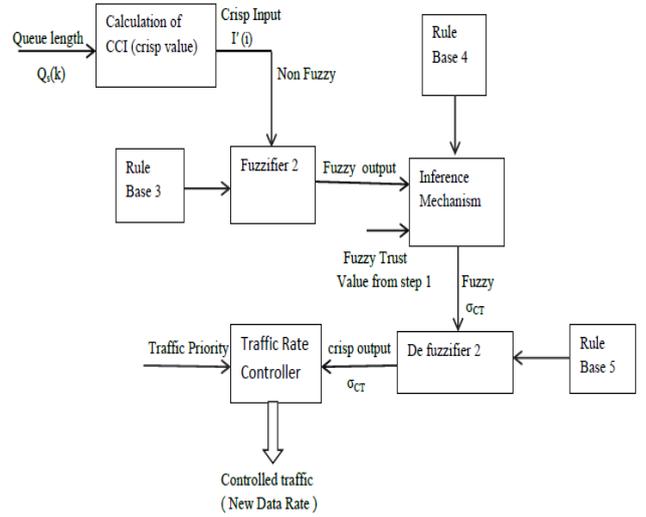

Fig. 3  Flow diagram of TFCC (step two)

The total CCI for $i^{th}$ node is given by:

$$I'(i) = \left( \prod_{k=1}^{N_i+1} I_k'(i) \right)^{\frac{1}{(N_i+1)}}$$

Mathematically,

$I_k'(i) = 1 - I_k(i)$  where

$I_k(i) = \epsilon$  for  $Q_s(k) \leq C_{TH}(Min)$

$I_K(i) = 1$  for  $Q_s(k) > C_{TH}(Max)$  and

$I_k(i) = (1-\epsilon)\left( \dfrac{Q_s(k) - C_{TH}(\min)}{C_{TH}(\max) - C_{TH}(\min)} \right) + \epsilon$

for  $C_{TH}(Min) \leq Q_s(k) < C_{TH}(M_{AX})$  &  $\epsilon$  is a small quantity less than one.

$I'(i)$ for $i^{th}$ node is taken as the crisp input variable of the Fuzzifier 2. The Fuzzified CCI value of the trusted sensor nodes are obtained at the output of Fuzzifier 2 using rule base 3 as shown in Table IV.

TABLE IV
Rule Base 3

| Crisp Input range of I'(i) | Fuzzy variable range of I'(i) at the output of Fuzzifier 1 |
|---|---|
| 0 – 0.3 | VLC ( Very Low Congestion ) |
| 0.25 – 0.55 | LC ( Low Congestion ) |
| 0.5 – 0.75 | MC ( Medium Congestion ) |
| 0.7 – 1 | HC ( High Congestion ) |

### 2) Computation of CCI – Trust Metric, $\sigma_{CT}$ :

The fuzzy Trust value and I′(i) of the sensor nodes are mapped to the CCI-Trust metric, Fuzzy $\sigma_{CT}$ at the inference mechanism by rule base 4 as shown in Table V.

TABLE V
Rule Base 4

| Fuzzy I′(i) | Fuzzy Trust | Fuzzy $\sigma_{CT}$ |
|---|---|---|
| VLC | MT/HT | VL (Very Low) |
| LC | MT /HT | L (Low) |
| MC | MT | M (Medium) |
| HC | MT | H (High) |
| MC | HT | M (Medium) |
| HC | HT | H (High) |

The Fuzzy Trust value of VLT and LT implies the misbehavior of the malicious nodes and hence is not considered in rule base 4. The crisp value of $\sigma_{CT}$ is obtained at the output of Defuzzifier 2 which is controlled by the rule base 5, as depicted in Table IV below.

TABLE IV
Rule Base 5

| Fuzzy variable $\sigma_{CT}$ | Crisp output range of $\sigma_{CT}$ |
|---|---|
| VL | 0-0.2 |
| L | 0.15 – 0.5 |
| M | 0.45 – 0.8 |
| H | 0.75 – 1.0 |

The parameter $\sigma_{CT}$ of each benevolent node, thus obtained from Defuzzifier 2, reflects the condition of the node in terms of its trust value and congestion status. The high value of $\sigma_{CT}$ implies highly congested node.

### 3) Congestion Control and Rate Adjustment :

The congestion control refers to the mechanism and techniques for controlling congestion in order to keep the offered load below the capacity of the network. The existing trust based routing protocols mostly select nodes with high trust value. Hence, in most of the cases, trusted nodes are highly congested. In our protocol, the congestion of the trusted nodes are controlled at the Traffic Rate Controller (TRC) which adjust the rate of the traffic flow on the basis of $\sigma_{CT}$ value and the priority of the corresponding traffic. The local traffic source transmit packet with highest priority. The priority of the other traffic from child nodes are assigned as per the trust value of the node with respect to its parent node.

For example, let us consider k1, k2 and k3 are the child nodes of the parent $i^{th}$ node as shown in Fig. 4. All the child nodes would like to transmit data packets towards its parent node. Let the trust value of k1, k2 and k3 nodes with respect to $i^{th}$ node be $T_{k1i}$, $T_{k2i}$ and $T_{k3i}$ respectively. In TFCC, the local traffic generated from $i^{th}$ node is transmitted with highest priority. The priority of the remaining traffic from k1, k2 and k3 nodes are assigned on the basis of the trust value of the nodes. Hence, if $T_{k1i} > T_{k2i} > T_{k3i}$, then, the priority of the traffic generated from $K_1$ node is higher than that from $K_2$ node. Similarly, the priority of the traffic from $K_2$ node is higher than that from $K_3$ node and so on.

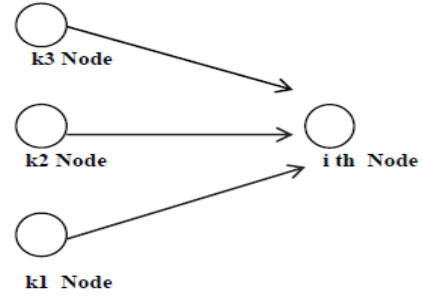

Fig. 4 Data packet transmission from child to parent node

The rate of data packet transmission is adjusted at TRC on the basis of the traffic priority and $\sigma_{CT}$ as shown in Fig 3. The high priority traffic is transmitted at higher rate than the low priority traffic until the condition $C_{TH}(Min) \leq Q_s(k) < C_{TH}(M_{AX})$ is satisfied at the node. When $Q_s(k) > C_{TH}(M_{AX})$, the rate of the data packet transmission is reduced accordingly to maintain the optimum condition. Thus controlled traffic flow is obtained at the output of TRC. So, in TFCC model the congestion condition of the trusted nodes are adjusted which in turn improves the network congestion as well as the throughput of the network.

## IV. SIMULATION RESULTS

In this section, we investigate the merits of our protocol through some simulation experiments. We have considered an arbitrary network comprising of 100 nodes deployed randomly into a field of dimension 50 m * 50 m. The set up under consideration has approximately 50% benevolent nodes while the rest are deemed malicious. The malicious nodes are blocked against data packet routing. All the links considered are duplex links. The different paths connecting each individual benevolent node deployed into the field and the source to sink are obtained via Link State Routing Protocol (LSRP).

We have compared our algorithm with the existing protocol QCCP-PS [2], plotted in blue, and FCCTF [12], plotted in black, while our graph is plotted in red, as shown in Fig. 5. The normalized throughput is taken as a metric for comparing the protocols. Time is plotted along x-axis while normalized throughput is plotted along the y-axis. The throughput of the sensor network is defined as the number of the packets passing through the network in unit time. From the simulation experiment it is observed that the proposed TFCC algorithm gives better result compared to the other two existing protocols.

In QCCP-PS, congestion control is proposed by using Rate Adjustment Unit (RAU) but it does not consider the practical conditions pertaining to a sensor network which deteriorate with time such as node energy, insurgence of malicious packets, nodes turning malicious in nature, trustworthiness of nodes etc. As TFCC takes all these into consideration, its performance is far better than QCCP-PS in a holistic sense. The other algorithm FCCTF consider both trust and congestion simultaneously but trust computation in FCCTF is less scientific as it does not consider the node energy. The remaining energy of the node is taken as one of the trust metric in proposed TFCC algorithm showing better result compared to FCCTF. Furthermore, in TFCC algorithm LSRP is applied on the trusted links of the benevolent nodes for getting available routes from source to sink .

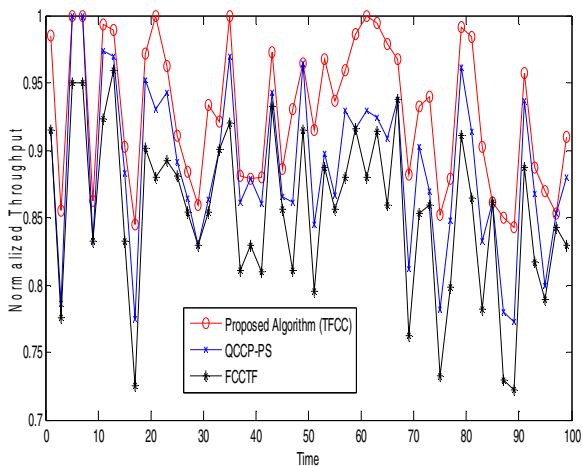

Fig. 5   Time versus normalized throughput graph

## V. Conclusions

In this paper, we have presented a new trust based congestion control protocol for WSNs and WMSNs, in which the rate of traffic flow is dynamically adjusted on the basis of the priority of the traffic. It is very much relevant, as a recent research topic since multimedia applications in WSNs is gaining popularity nowadays. Our protocol considers the impact of the network congestion due to the misbehavior of the faulty nodes and thereby minimizes their effect during packet transmission. TFCC protocol resolves congestion of the individual nodes and hence reduces network congestion significantly. The congestion control and quality of service are two closely bounded issues and improving one implies improving the other. The simulation and experimental results indicate that TFCC provides higher throughput compared to other similar protocols and thereby outperform its peers. However, we have adopted an interval type I fuzzy set for computational purposes. In spite of its several advantages, fuzzy sets are limited by the fact that the degree of membership of an entity in a particular class also has a measure of uncertainty associated with it. Hence, this leads to a certain degree of uncertainity which is associated with the resulting data packet transmission rate to the various nodes. This can be overcome by means of either type II fuzzy computation or ordinary deterministic non-fuzzy computation. Moreover, the proposed TFCC protocol is tested only on a small network architecture comprising of 100 nodes in a field of area 50 m * 50 m. So, we need to test its adaptability to larger networks consisting of large number of nodes spreaded over larger areas. In future, we will also analyze the total congestion throughout the route in upstream direction from source to sink.